\documentclass[twocolumn,showpacs,amsmath,amssymb,aps,prl]{revtex4}

\usepackage{graphicx}
\usepackage{bm} 
\usepackage{epsfig}

\begin{document}

\title{Rocking 
 ratchets in 2D Josephson  networks: collective effects and current reversal}

\author{Ver\'onica I. Marconi}

\affiliation{Institut de physique, Universit\'e de Neuch\^{a}tel,
 CH-2000 Neuch\^{a}tel, Switzerland.}

\begin{abstract}

 A  detailed numerical study on the  directed motion of ac-driven  vortices and antivortices in  2D
Josephson  junction arrays (JJA) with an asymmetric  periodic pinning potential  is reported.  Dc-voltage
rectification shows a strong dependence on vortex density as well as an inversion of the vortex flow
direction with ac amplitude for a wide range of  vortex density around $f$=1/2 ( $f$=$Ha^2 / \Phi_0$), in
good agreement with recent  experiments by Shal\'om and Pastoriza [Phys. Rev. Lett. {\bf 94},  177001 (2005)]. The study
of vortex structures, spatial and temporal correlations, and vortex-antivortex  pairs formation gives
insight into a purely collective mechanism behind the current reversal effect.
\end{abstract}
\maketitle

Ratchet systems  constitute a vast  subject of  present interest, extensively studied from biology
 to a wide spectrum of  fields in physics \cite{julicher, reimann, menon, astumian, linke}. Very
recently, promising applications in micro-nano technology have motivated   a huge amount of  additional 
works, mainly regarding the controlled motion of nanoparticles, colloids, electrons or magnetic flux quanta.
The main characteristic of these systems is their ability to transport particles with a non-zero drift
velocity in the absence of any net macroscopic force. The simplest realization of this idea  is  the
rectification of  an ac-driven  single particle motion in an asymmetric periodic potential or ratchet
potential,  known as a {\it rocking ratchet}.  Ratchet effect in superconductors (SC) has been, in 
particular, an attractive subject of study since the proposal of using it to remove  undesirable trapped
vortices producing noise in SC devices \cite{barabasi99}. Notably, several designs of asymmetric pinning
potentials for the motion  of flux quanta have been recently realized, such as dots-antidots arrays of 
pinning sites in SC films \cite{films}.  Even more, ratchet designs without spatial asymmetry have  been
successfully achieved in layered SC \cite{cole}.   Small Josephson junction (JJ) systems were investigated
as well:  SQUIDs \cite{sols},   1D ladders\cite{falo, trias}, long JJ \cite{goldobin} and  quasi-1D quantum
JJA \cite{majer}.   In particular, on classical  JJA, the first  design of a device for kinks motion
rectification was proposed \cite{falo} and measured \cite{trias} in 1D.    

Large 2D JJA, which are excellent systems to study statistical mechanics experimentally \cite{tink} and
where collective effects could play a relevant role,  had not been studied until very recently.
Rectification experiments on   2D JJA  were reported by Shal\'om and  Pastoriza  \cite{pastoriza}.  They
modulated the  distance  between the SC islands  thus generating  a ratchet potential for vortex motion.
Dc-voltage rectification was then clearly observed and its intriguing  features, such as a  non trivial
dependence on both, the applied ac current and  vortex density, and an inversion of the vortex flow
direction, prompt further investigation. The aim of this  work is to interpret  such experiments  and
understand  the role of collective effects.  We show numerically that the  overdamped RSJ model for  large
2D JJA with asymmetrically  modulated critical currents, reproduces successfully  the experiments
\cite{pastoriza} and allows for a  detailed characterization  of the device useful  for performance
enhancement and future experiments.  In addition,  a complete  study of the  vortex dynamics gives insight
into a non trivial collective mechanism behind the current reversal (CR) effect.

 We study the overdamped vortex dynamics in  square {\it 2D asymmetrically modulated}  JJA using the
resistively shunted junction (RSJ) model \cite{dyna} and solving numerically the superconducting phase
dynamics.   The current flowing in the junction between two superconducting islands  is the sum of the
Josephson supercurrent and the normal current: 
$
I_{\mu }({\bf n})=I_{0}
\sin \theta_{\mu }({\bf n})+\frac{\Phi_{0}}
{2\pi cR_{N}}\frac{\partial \theta _{\mu }({\bf n})}{\partial t}+
\eta _{\mu}({\bf  n},t),
$
 where  $I_0$ is  the critical current of the junction between
the two superconducting islands sited in ${\bf n}$ and ${\bf n}+{\bf \mu}$ on a square lattice 
[${\bf n}$=$(n_x,n_y)$, ${\bf \mu}$=${\bf \hat x}, 
{\bf \hat y}$],  $\theta_{\mu}({\bf n})$=$\theta({\bf n}+{\bf \mu})-\theta({\bf
n})-A_{\mu}({\bf n})$  the
gauge invariant phase difference with $A_{\mu}({\bf n})$=$\frac{2\pi}{\Phi_0}
\int_{{\bf n}a}^{({\bf n}+{\bf\mu})a}{\bf A}\cdot d{\bf l}$,  
$\Phi _{0}$ the flux quantum, $a$ the lattice parameter, 
  and $R_N$ the normal state resistance.
  Under a magnetic field  $H$, we have
$\Delta_{\mu}\times A_{\mu}({\bf n})$=
$2\pi f$,  $f$=$H a^2/\Phi_0$ being  the   frustration parameter,
a measure of the vortex density.
  The Langevin
noise term $\eta _{\mu }$  models the contact with a thermal bath at temperature
$T$ and satisfies $ \langle \eta _{\mu }({\bf n},t)\eta _{\mu ^{\prime }} ({\bf
n^{\prime }},t^{\prime })\rangle$=$\frac{2k_BT}{R_{N}}\delta _{\mu ,\mu ^{\prime
}}\delta _{ {\bf n},{\bf n^{\prime }}}\delta (t-t^{\prime })$. 
The ratchet potential for vortices is introduced  by  modulating the critical currents $I_{0}$=$E_J\Phi_{0}/
2\pi$, $E_J$ being the Josephson coupling energy  between islands.  We consider  a sawtooth modulation 
$I_0(n_x)$ increasing linearly from  
$I_{0_{min}}$  to $I_{0_{max}}$ within each period, $p$, as an approximation to the experimental modulation. The ratchet potential amplitude is then measured  by $\Delta U \propto \Delta I_0 $=$ (I_{0_{max}}-I_{0_{min}})$.
The system is  driven by $I_{ext}$=$I_{ac}\sin(2\pi\omega_{ac}t)$ in the ${\bf \hat y}$ direction. Periodic
boundary  conditions in both directions are taken (see model details in  \cite{prb2}). The dynamical equations
for the superconducting phases are obtained after considering  conservation of the current in each node. The
resulting set of  Langevin  equations are solved using  a $2^{nd}$ order Runge-Kutta algorithm with time
step $\Delta t$=$0.1\tau_J$ ($\tau_J$=$2\pi cR_N \overline{I_0} /\Phi_0$, where $\overline{I_0}$ is the
averaged critical current) and  integration time    $2.10^6 \tau_J$   after the equilibration transient.  
We calculate:  
 mean dc voltage as $\langle V_{dc} \rangle$=-$\langle d\theta_y({\bf n})/dt\rangle$ \cite{prb2} vs both $I_{ac}$ and $f$ where $<$..$>$ stands for both time and space average;  vorticity  at the plaquette ${\bf \tilde n}$=$(n_x+1/2,n_y+1/2)$,  as $b({\bf \tilde n})$=$-\Delta_\mu\times{\rm nint}[\theta_\mu({\bf n})/2\pi]$ \cite{ijja} (nint[] is the nearest integer function).   Currents are normalized by   $\overline{I_0}$, voltages by $R_
{N}\overline{I_0}$, temperature by $\overline{E_J}/k_B$ and frequencies by
$(\tau_J)^{-1}$.  
Most of the results  shown  are  (except indication) for
 $p$=$8$,  $I_{0_{min}}$=$0.5
\overline{I_0}$, $I_{0_{max}}$=$1.5 \overline{I_0}$, $\omega_{ac}$=$10^{-4}$,
 $T$=$0.05$ and 32$\times$32 arrays.
 Other parameters as $p$=$7,10,15$ and $\Delta I_0$=$0.5,1.5,2$ were as well investigated.  The full range of  magnetic field was explored,
from  a single vortex  to the fully frustrated case ($f$=$1/2$) in
 system with  $L$=$L_x$=$L_y$=$32,64,100,128$.
\begin{figure} 
\centerline{\includegraphics*[width=7.cm]{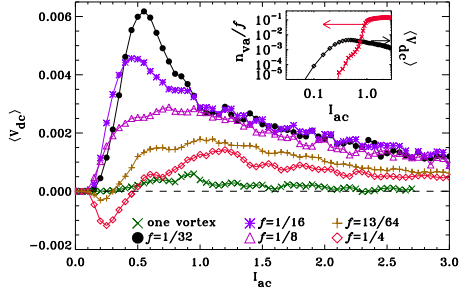}}
\caption{
Rectified vortex motion:  
$I_{ac}-V_{dc}$ characteristics examples increasing
 vortex density, from one vortex in the sample ($f$=$1/L^2$) towards $f$=$1/2$. 
Inset: $\langle V_{dc} \rangle$ vs $I_{ac}$ for $f$=$1/32$, in log-log scale (right y-axis)
 and vortex-antivortex pairs density normalized by frustration,
 $n_{va}/f$ (left y-axis). 
}
\label{fig:ivs}
\end{figure}

{\it Results}.  In Fig.\ref{fig:ivs}  we show    mean dc-voltage, $\langle V_{dc} \rangle$, vs the amplitude
of the ac current, $I_{ac}$, for different vortex densities, $f$.  These results imply rectification: a net
directional vortex motion  parallel to the ratchet modulation in response to an alternating force. For low
$f$, $\langle V_{dc} \rangle$ is maximum at an optimal value of $I_{ac}$ and decreases slowly for larger
$I_{ac}$ as the ratchet potential is effectively washed out by the fast vortex motion. This  behavior is
qualitatively the one expected for a single particle rocking ratchet system \cite{reimann}. Let us note 
however that even for highly diluted vortex systems the intensity of rectification is not additive:  for
example with $f$=$1/32$ in a $32 \times 32$ array we have 32 vortices but as we can see in Fig.\ref{fig:ivs}
the response is  $\sim$12 times, and not 32 times the response of one single vortex ($f$=$1/L^2$). This
non-additivity is even more pronounced for higher $f$. By increasing frustration from $f$=$1/32$ to
$f$=$1/2$ the rectification peak presents both a reduction and a shift of the optimal $I_{ac}$ value. A more
intriguing feature appears when $f$ is further increased.  For $0.17  \lesssim f < 1/2$ an inversion of the
vortex flow direction is observed at small $I_{ac}$ values and  the window with $\langle V_{dc} \rangle <
0$  becomes wider as $f$ is increased.    The non monotonic  behavior observed as a function of $f$   is a clear evidence that
vortex-vortex  interactions  are relevant for the rectification effect at all $f$.   On the other hand it is well known that the frustrated XY-model, equivalent to our model for  2D JJA 
under an external $H$ \cite{tink}, can  also present   a finite density of thermally excited
vortex-antivortex pairs, $n_{va}$=$(\langle|b({\bf\tilde{n}})| \rangle-f)/2$, over the magnetic field
induced vortex pattern. It is then important to under\-stand its contribution to the ratchet effect. In  the
inset of Fig.1 we plot both $\langle V_{dc}\rangle$ and $n_{va}/f$ vs $I_{ac}$ for $f$=$1/32$.  We see that
$n_{va}$  increases monotonically and saturates to $10\%$ of $f$ at $I_{ac} \sim 0.8$. We find the same
behavior  for all $f$ but the saturation value of $n_{va}$ decreases with increasing $f$, being less than 
$0.01f$ for $f>1/8$. We conclude that the effect of pairs  is  almost negligible in the  CR regime, but it
is  relevant for diluted vortex lattices  at large applied currents and  they must be considered  to
reproduce properly the experimental rectification tails. 
\begin{figure}
\centerline{\includegraphics*[width=7.2cm]{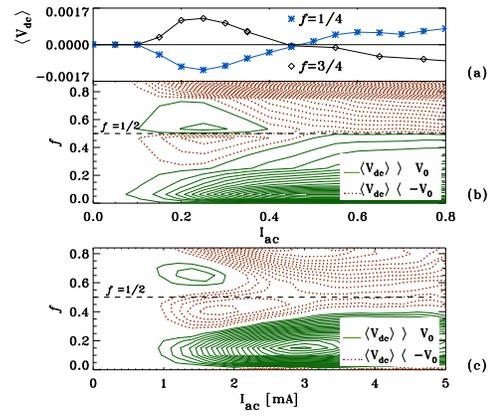}}
\caption{
 (a) $\langle V_{dc} \rangle$ vs $I_{ac}$ for  $f$=$1/4$ and $f$=3/4.  (b) Numerical $\langle V_{dc} \rangle$ contour plots in $I_{ac}-f$ plane 
 at $T$=$0.05E_J/k_B$ and
$\omega_{ac}$=$10^{-4}(1/\tau_J)$.  Red dotted
lines represent $\langle V_{dc}\rangle < -V_0$ and green solid lines, $\langle V_{dc}
\rangle >  V_0$=$5.10^{-5} R_NI_0$.
(c)
Experimental  contours at T$=$3.8K, $\omega_{ac}$=$1 kHz$ and  $V_0$=$15 nV$ (see  \cite{pastoriza}).
}
\label{fig:cont}
\end{figure}

A summary of our numerical results on voltage rectification vs  vortex density is presented  in 
Fig.\ref{fig:cont} in comparison with experiments \cite{pastoriza}.  In Fig.\ref{fig:cont}(a) we show  
numerical curves for $f$=$1/4$ and $f$=1$-f$=3/4.  From the symmetries of the frustrated XY Hamiltonian:  
${\cal H}(f)$=${\cal H}(-f)$=${\cal H}(1-f)$ \cite{tink},  we expect symmetrical transport properties around
$f$=$1/2$.  Since vortex and antivortex motion is   rectified in the same direction   we have  $\langle
V_{dc} \rangle(f)$=$-\langle V_{dc} \rangle$(1$-f$). This means that in the 0 $\leq f \leq$ 1/2 range of
magnetic field the dissipation arises from the motion of ``positive'' vortices while the  flux of
``negative''  vortices (antivortices) yields  dissipation in the 1/2 $\leq$ $f$ $\leq$ 1  range.  In
Fig.\ref{fig:cont}(b) contour plots of $\langle V_{dc}\rangle$  show the occurrence of voltage
rectification  in the $I_{ac}-f$ plane.  Current reversals occur only in the regime of strong vortex
repulsion around $f$=$1/2$, in a  single interval,   $0.17 \lesssim f < 0.5$.  Comparing
Fig.\ref{fig:cont}(b-c) we see  a good  qualitative agreement  with experiments.
\begin{figure} 
\centerline{\includegraphics*[width=8.0cm]{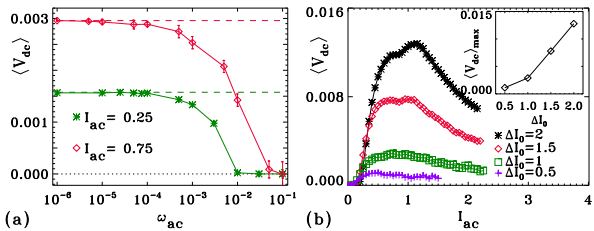}}
\caption{ Rectification dependences for $f$=$1/8$: (a){\it On $\omega_{ac}$:} 
at $I_{ac}$=$0.25 (\star)$ and
$I_{ac}$=$0.75 (\diamond)$  ($\langle V_{dc} \rangle_{max}$ , see Fig.1). (b){\it On pinning potential:}  $\langle V_{dc}\rangle$-$I_{ac}$
  vs  $\Delta I_0$.  Inset:  $\langle V_{dc}\rangle_{max}$ vs $\Delta I_0$. }
\label{fig:vs}
\end{figure}
 
 For applications it is also important 
 to  accurately determine the range of parameters where rectification would be observable.
  In  Fig.\ref{fig:vs}(a)  we see a well defined  saturation  plateau for $\langle V_{dc}\rangle$
 at low frequencies. At high $\omega_{ac}$ the fast changes in 
the potential slope do not allow vortices to adapt to   the ratchet potential 
 during a  cycle. For all  $I_{ac}$ the mean voltage thus
 decays, becoming negligible at 
 $\omega_{ac}$$\gtrsim$0.01$q(\tau_J)^{-1}$$\sim$1$/\tau_{rel}$.
 Then we  predict
 an observable rectification up to $\sim$100kHz 
in good agreement with \cite{pastoriza}, if we  use typical parameters,  $R_N$=$0.01 \Omega$ and $I_0$=$1 \mu A$,
to estimate our unit of frequency $(\tau_J)^{-1}\sim$30MHz.
  Rectifica\-tion can also  depend  on the parameters
 of the ratchet potential.  In Fig.\ref{fig:vs}(b)  we see that $\langle V_{dc} \rangle$ monotonically increase with $\Delta I_0$ for all $I_{ac}$, and the optimal rectification increase approximatively linearly  for $1.5 \lesssim \Delta I_0 \lesssim 2$ (see inset) and for $ \Delta I_0 \sim 2$ inversion disappears  \cite{fut}. We thus see that the ratchet effect could be experimentally improved up to one order of magnitude by modulating accordingly the Josephson couplings.
On the other hand we find that rectification has no appreciable changes for the experimental values $p$=7,10,15 nor sharp commensurability effects between $f$ and $p$. 
We    also analyze the range of temperature where rectification occurs and size effects.
We find  a  wide range of applicability,  $0 \le T \lesssim 0.4$, what makes 
 JJA advantageous and that the main characteristics of our   results are almost size  independent  for  $L\,>\,32$ \cite{fut}.

\begin{figure}
\centerline{\includegraphics*[height=7.cm]{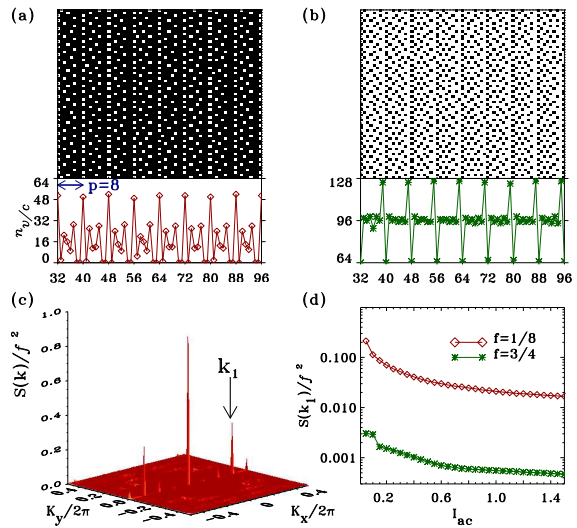}}
\caption{Vortex structures: static $b({\bf \tilde n}$) obtained quenching
 from high $T$
towards $T$=$0$ ($L$=$128$ but a fraction  is
shown for transparency. Vortices are white plaquettes). In (a)  $f$=$1/8$  and (b) $f$=$3/4$, both with  $\langle n_{v/c}\rangle$ below, the number of
vortices per column.  (c) 
$S(\bf { k } )/\it{f}^2$ corresponding to $b({\bf\tilde{n}})$ in  (a) with  clear peaks due to the ratchet potential,  as $k_1$.
 (d) $S(k_1)/f^2$  vs $I_{ac}$ for $f$=$1/8, 3/4$  at
$T$=$0.05$ and $\omega_{ac}$=$10^{-4}$.
}
\label{fig:sf}
\end{figure}

In order to understand collective transport effects and in particular {\it current reversal}, we  study vortex structures and spatial and temporal correlations. 
 In Fig.\ref{fig:sf}(a-b) we show   vorticity snapshots for  $f$=$1/8,3/4$ at $I_{ac}=0$ illustrating the competition between the ratchet potential, which induces a modulated vortex density, and the repulsive vortex-vortex interaction that favors a uniform vortex density. For all $f$ we find that vortices organize in dense vortex walls strongly pinned at the  sawtooth minima. Interestingly,  when CR starts, the static vortex structure present an   averaged number of vortices in the walls similar to a checkerboard pattern, i.e. $\langle n_{v/wall}\rangle$$\sim$$L_y/2$.
The degree of periodic order is reflected by the intensity of the peaks
 at ${\bf k}$=$(n 2\pi/p, 0)$ of the averaged vortex structure factor 
$S({\bf k})$=$\left\langle\left|\frac{1}{L^2}\sum_{\bf \tilde n}
b({\bf \tilde n})\exp(i{\bf k}\cdot{\bf \tilde n})\right|^2\right\rangle$
shown in Fig.\ref{fig:sf}(c). A ring appears around the central peak   
representing the  disordered  structure observed in the inter-wall region, inside which the vortex density becomes more uniform as $f$ approaches $f$=$1/2$ (compare $\langle n_{v/c}\rangle$ in  (a) vs (b)).  
 When the system is driven the  ratchet induced periodicity survives but its amplitude decays with increasing $I_{ac}$. We find that
 the degree of order decreases considerably for $f$ around $f$=$1/2$ in the  CR regime, although vortex walls at small $I_{ac}$ are still well defined, see Fig.\ref{fig:sf}(d).  
We now  analyze  the instantaneous vortex dynamics.
\begin{figure}
\centerline{\includegraphics*[width=7.5cm]{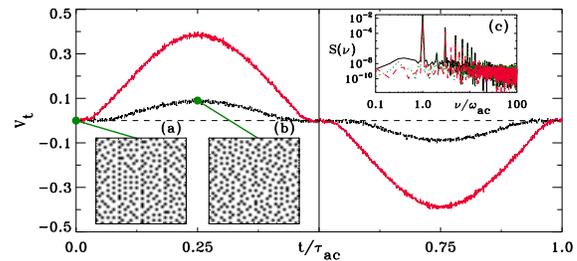}}
\caption{Time voltage response  to an ac-drive of $I_{ac}$=0.3 for $f$=$3/4$
(smaller amplitude curve, black line),  with  
   $\langle V_{dc} \rangle$$>$0, and  $I_{ac}$=0.7 (larger  amplitude, red line), with $\langle V_{dc} \rangle$$<$0. Insets (a),(b): instantaneous vorticity  for $I_{ac}$=0.3   and $t/\tau_{ac}$=$0, 0.25$ respectively.
 (c) Power spectrum  for $f$=3/4 (solid black line),
$f$=$1/8$ (dotted green line) and $f$=$1/16$ (dashed red line).
}
\label{fig:dyn}
\end{figure}
  In Fig.\ref{fig:dyn} we show examples of typical  voltage response vs time during a period at which its steady value $\langle V_{dc} \rangle$ (see Fig.\ref{fig:cont}(a)) is already achieved.  The amplitude of  the almost symmetrical   temporal oscillations 
 around $V_{t}$=0 is much larger, one order of magnitude, than $\langle V_{dc} \rangle$ for all $I_{ac}$. This implies that rectification arises from  tiny differences between the two directions of motion
and explains  the low efficiency of the device  and the  need of  very large $\tau_{ac}$=$\omega_{ac}^{-1}$ to resolve the effect. The dc-limit analysis confirms this result   \cite{fut}.  
Insets (a),(b) in Fig.\ref{fig:dyn} show $b({\bf \tilde n})$  snapshots   at time  [0,0.25]$\tau_{ac}$ respectively. Most of the time the moving vortex lattice is highly disordered 
(as seen at $t/\tau_{ac}$=$0.25$), except for small forces close to  depinning (as in $t/\tau_{ac}$$\sim$0)
where the ratchet-induced periodicity is more visible.
  This fluctuating order produces the small peaks in the averaged structure factor  showed in Fig.\ref{fig:sf}(d).  
 In Fig.\ref{fig:dyn}(c) voltage power spectrum for different vortex densities are analyzed. Time correlations appear  clearly at $\omega_{ac}$ and  higher  harmonics, as it is expected for one particle in a ratchet potential. Note however that interactions  modify the harmonics structure, as it is seen  comparing $f$=3/4 with more diluted vortex lattices, $f$=$1/8$ and $f$=$1/16$  where more harmonic peaks appear. We thus see that temporal order is only determined by the ac-drive as it is expected for vortices  moving  incoherently. No trace of a washboard frequency is observed for any $f$ nor $I_{ac}$.  Therefore no dynamical phase transitions are induced by the ac-drive and the system behaves as a modulated driven plasma.
 
{\it Discussion}.
Our results present some similarities but also some marked differences with other 2D vortex ratchet  systems. First, let us note that inverted rectification appears {\it only at low drives within a single range of high $f$  around $f$=1/2}  (Fig.\ref{fig:cont}). This is similar to CR observed in arrays of  dots in SC films by Villegas  {\it et al.} \cite{films}, but their explanation based on interstitial vortices moving over an inverted ratchet potential,  is not suitable for JJA as interstitial can not be defined. Our results are also different to the  multiple  CRs observed and explained by de Souza Silva {\it et al.} in  SC films with antidots arrays  \cite{films}. The differences can be attributed to the strong pinning generated by dots or antidots, which can localize  a few vortices in very small regions and produce a very coherent individual motion. In  our case an extended weak potential pins and deforms collectively the vortex density although individual vortex motion remains highly incoherent. In addition, underdamped dynamics was also shown to provide CR \cite{films}. In our case no evidence of an effective dynamically induced
inertia is found however, as no hysteresis in  Fig.\ref{fig:ivs},  nor out of phase voltage response is observed, Fig.\ref{fig:dyn}. 
CR in  JJA  thus presents novel features deserving discussion.

Since CR is observed near the onset of dissipation, an analysis of the asymmetric depinning is useful. 
In the absence of vortex interactions it is clear that the critical current in the easy direction of the ratchet potential, $I_{dep}^{easy} \sim\Delta U/ (p-1)$ is smaller than the one in the hard direction 
$I_{dep}^{hard} \sim\Delta U$. Therefore the onset of rectifica\-tion occurs when $I_{ac} \sim I_{dep}^{easy}$ with a net motion in the easy direction. In presence of interactions the situation can be rather different \cite{films}, since vortices tend to screen the ratchet potential thus modifying the critical currents and mobility of single vortex excitations in each direction differently.  It is clear that a  small tilt of the ratchet poten\-tial in the hard direction produces well defined walls with a high vortex density at the local minima of the sawtooth. On the contrary a tilt in the easy direction favors a smoother vortex density profile.
Calculating the  force exerted on a vortex by a periodic array of vortex walls we can easily show that
  $I_{dep}^{hard}$ is reduced   by an amount  proportional to the vortex density  at the walls,  $\tilde{I}_{dep}^{hard}$=$\Delta U - \alpha f_{wall}$, while in the easy direction  ${I}_{dep}^{easy}$ remains unchanged.
Hence, CR can occur above a critical vortex density, $f_c$, such that  $\tilde{I}_{dep}^{hard}$=$ {I}_{dep}^{easy}$ at $f_c$, yielding $f_c \sim  \frac{\Delta U}{\alpha} ( \frac{p-2}{p-1})$. This simple argument is consistent with the fact that inversion is observed at high $f$, disappears with increasing $\Delta U$ and is weakly dependent on $p$ (details and dc limit confirming this scenario in \cite{fut}).
At larger $I_{ac}$ the  dynamical smoothening 
of the vortex density reduces  the screening  and  the easy direction for rectification is   recovered. Fig.\ref{fig:ivs} shows that  this recovering occurs at higher $I_{ac}$ for higher  $f$, consistent with the existence of denser walls at high $f$.
 Our results thus show that vortex-vortex interactions can overscreen the asymmetry of the ratchet potential as  was noted in \cite{films}
 but the mechanism is {\it purely collective}, since individual\- vortex motion is highly incoherent like in a driven modulated fluid.  Indeed, this effect  could be  relevant for  a wide family of interacting  many-body ratchet systems  \cite{fut}. 

I acknowledge useful discussions with P. Martinoli, A.B. Kolton,  D. Dom\'{\i}nguez, D.E
Shal\'om (also  for providing  experimental data), H. Pastoriza and H. Beck. This work was supported by the Swiss NSF.


\end{document}